\documentclass{article}
\usepackage{amsfonts}
\usepackage{amsmath}
\usepackage{graphicx}

\newtheorem{theorem}{Theorem}

\newtheorem{corollary}{Corollary}

\newenvironment{proof}[1][Proof]{\noindent\textbf{#1.} }{\ \rule{0.5em}{0.5em}}
\input{tcilatex}
\begin{document}

\title{{\Large {\textbf{Distortion risk measures for sums of dependent losses%
}}}}
\author{{\large \textbf{Brahim Brahimi, Djamel Meraghni}, \textbf{and
Abdelhakim Necir}} \\
{\small {\textit{Laboratory of Applied Mathematics, Mohamed Khider
University of Biskra, }}}\\
{\small {\textit{Biskra 07000, Algeria}}\smallskip }}
\date{{\small 29 June 2010}}
\maketitle

\begin{abstract}
\noindent We discuss two distinct approaches, for distorting risk measures
of sums of dependent random variables, which preserve the property of
coherence. The first, based on distorted expectations, operates on the
survival function of the sum. The second, simultaneously applies the
distortion on the survival function of the sum and the dependence structure
of risks, represented by copulas. Our goal is to propose risk measures that
take into account the fluctuations of losses and possible correlations
between risk components.\medskip

\noindent \textbf{R\'{e}sum\'{e}.} Nous discutons deux approches distinctes,
de distortion des mesures de risque de la somme de variables al\'{e}atoires d%
\'{e}pendantes, qui conservent la propri\'{e}t\'{e} de coh\'{e}rence. La
premi\`{e}re, bas\'{e}e sur les esp\'{e}rances distordues, agit sur la
fonction de survie de la somme. La seconde, applique des d\'{e}formations
simultan\'{e}es sur la fonction de survie de la somme et sur la structure de
d\'{e}pendance des risques, repr\'{e}sent\'{e}e par une copule. Notre
objectif est de proposer des mesures qui prennent en compte les fluctuations
des pertes et des corr\'{e}lations \'{e}ventuelles entre les composantes
d'un risque multivari\'{e}.\bigskip

\noindent \textbf{MSC 2010:} 60B05, 62H20, 91B30. \medskip

\noindent \textbf{Keywords:} Coherence; Dependence structure; Distortion
function; Risk measure; Risk theory; insurance; Wang transform.
\end{abstract}

\section{\textbf{Introduction}}

\noindent Risk measures are used to quantify insurance losses and financial
assessments. Several risk measures have been proposed in actuarial science
literature, namely: the Value-at-Risk (VaR), the expected shortfall or the
conditional tail expectation (CTE), and the distorted risk measures (DRM).
Before introducing and interpreting the DRM, it is necessary to fix a
convention of profit and loss appropriate to the application to the market
finance, the credit risk and to the insurance. Let $X$ be a random variable
(rv), representing losses (or gains) of a company, with a continuous
distribution function (df) $F.$ The DRM of rv $X,$ due to Wang \cite{Wan95},
is defined as follows: 
\begin{equation}
\pi _{\psi }[X]:=\int_{0}^{\infty }\psi (1-F(x))dx,  \label{eq5.1}
\end{equation}%
where $\psi $ is a non-decreasing function, called distortion function,
satisfying $\psi (0)=0$ and $\psi (1)=1.$ In the actuarial literature the
following functions are frequently used:%
\begin{equation*}
\begin{array}{lcc}
\psi _{\rho }(s)=s^{\rho },\medskip & \text{for} & 0<\rho \leq 1,\medskip \\ 
\psi _{\kappa }(s)=\phi (\phi ^{-1}(s)+\kappa ),\medskip & \text{for} & 
0\leq \kappa <\infty ,\medskip \\ 
\psi _{\zeta }(s)=\min (s/(1-\zeta ),1)\medskip & \text{for} & 0\leq \zeta
<1, \\ 
\psi _{\alpha }(s)=s^{\alpha }(1-\alpha \ln s), & \text{for} & 0<\alpha \leq
1,%
\end{array}%
\end{equation*}%
where $\phi ^{-1}(u):=\inf \{x:\phi (x)\geq u\}$ is the quantile function of
the standard normal distribution $\phi .$ Constants $\rho ,\kappa ,\zeta $
and $\alpha $ are called distortion parameters. The functions $\psi _{\rho
}, $\ $\psi _{\kappa },$ $\psi _{\zeta }$ and $\psi _{\alpha }$ respectively
give rise to the so-called proportional hazard transform (PHT) (Wang \cite%
{Wan95}), the normal transform (Wang \cite{Wan00}), the CTE and the
look-back distortion (H\"{u}rlimann \cite{Hur98}). When $\rho =1$ and $%
\kappa =\zeta =0,$ there is no distortion and the corresponding DRM is equal
to the expectation of $X.$ For recent literature on risk measures one refers
to Denuit \textit{et al.} \cite{Den05} and Furman and Zitikis (\cite{Fur08a}%
, \cite{Fur08b}).\medskip

\noindent The problem of the axiomatic foundation of risk measures has
received much attention starting with the seminal paper of Artzner \textit{%
et al.} \cite{Art99}, where the definition of coherent risk measure was
first provided. A coherent risk measure is a real functional $\mu,$ defined
on a space of rv's, satisfying the following axioms:\medskip

\textbf{H1.} Boundedness from above by the maximum loss: $\mu(X)\leq \max
(X) .$

\textbf{H2.} Boundedness from below by the mean loss: $\mu ( X ) \geq\mathbb{%
E} (X).$

\textbf{H3.} Scalar additivity and multiplicativity : $\mu ( aX+b ) =a\mu (
X ) +b,$ for $a,b\geq0.$

\textbf{H4.} Subadditivity: $\mu(X+Y)\leq\mu ( X ) +\mu(Y).$\medskip

\noindent The only axiom that a DRM may lack in order to be a coherent risk
measure in the sense of Artzner \textit{et al.} \cite{Art99} is H4. However,
the subadditivity theorem of Choquet integrals (Denneberg \cite{Den94})
guarantees that $\mu(X+Y)\leq\mu(X)+\mu(Y)$ if and only if the distortion
function $\psi$ is concave. Hence, the DRM $\pi_{\psi} [ X ] $ defined in $( %
\ref{eq5.1} ) $ with a concave distortion $\psi$ is coherent. It is well
known that the CTE and the PHT are examples of concave distortion risk
measures, whereas the VaR is not. \medskip In traditional risk theory,
individual risks have been usually assumed to be independent. Traceability
for this assumption is very convenient, but not realistic. Recently in the
actuarial science, the study of the impact of dependence among risks has
become a major and flourishing topic. Several notions of dependence were
introduced to model the fact that larger values of one component of a
multivariate risk tend to be associated with larger values of the others. In
this paper, we deal with a vector of risk losses $\mathbf{X}=(X^{ ( 1 )
},...,X^{ ( d ) }),$ $d\geq2$ and we discuss the computation of the DRM of
the sum $Z$ of its components. When $X^{ ( 1 ) },...,X^{ ( d ) }$ are
independent and identically distributed, their sum is considered as a rv
whose df $G$ is the convolution of the marginal distributions of $\mathbf{X}%
. $ In this case, the DRM value of $Z,$ for a given distortion function $%
\psi $ may be obtained via formula $( \ref{eq5.1} ) ,$ that is%
\begin{equation}
\pi_{\psi} [ Z ] :=\int_{0}^{\infty}\psi ( 1-G ( z ) ) dz.  \label{eq5.2}
\end{equation}

\noindent Now, assume that $X^{ ( 1 ) },...,X^{ ( d ) }$ are dependent with
joint df $H$ and continuous margins $F_{i},$ $i=1,...,d.$ In this case, the
problem becomes different and its resolution requires more than the usual
background. Several authors discussed the DRM, when applied to sums of rv's,
against some classical dependency measures such as Person's $r,$ Spearman's $%
\rho$ and Kendall's $\tau,$ see for instance, Darkiewicz \textit{et al.} 
\cite{Dar04} and Burgert and R\"{u}schendorf \cite{Bur06}. Our contribution
is to introduce the copula notion to provide more flexibility to the DRM of
sums of rv's in terms of loss and dependence structure. For comprehensive
details on copulas one may consult the textbook of Nelsen \cite{Nel06}.
According to Sklar's Theorem (Sklar \cite{Skl59}), there exists a unique
copula $C: [ 0,1 ] ^{d} \rightarrow [ 0,1 ] $ such that%
\begin{equation}
H ( x_{1},...,x_{d} ) =C ( F_{1} ( x_{1} ) ,...,F_{d} ( x_{d} ) ) .
\label{eq5.3}
\end{equation}

\noindent The Copula $C$ is the joint df of rv's $U_{i}=F_{i} ( X^{ ( i ) }
) ,$ $i=1,...,d.$ It is defined on $[ 0,1 ] ^{d}$ by $C ( u_{1},...,u_{d} )
=H ( F_{1}^{-1} ( u_{1} ) ,...,F_{d}^{-1} ( u_{d} ) ) ,$ where $F_{i}^{-1}$
denotes the quantile function of $F_{i}.$ This means that the DRM of the sum
is a functional of both copula $C$ and margins $F_{i}.$ Therefore, one must
take into account the dependence structure and the behavior of margin tails.
These two aspects have an important influence when quantifying risks. If the
correlation factor is neglected, the calculation of the DRM follows formula $%
( \ref{eq5.2} ) ,$ which only focuses on distorting the tail. In order to
highlight the dependence structure, we add a distortion on the copula as
well. The notion of distorted copula has recently been considered by several
authors, see for instance Frees and Valdez \cite{Fre98}, Genest and Rivest 
\cite{Gen01}, Morillas \cite{Mor05}, Crane and van der Hoek \cite{Cra08} and
Valdez and Xiao \cite{Val10}. Given a copula $C$ and a non-decreasing
bijection $\Gamma: [ 0,1 ] \mathbb{\ \rightarrow} [ 0,1 ] ,$ the distorted
copula $C^{\Gamma}$ is defined by%
\begin{equation*}
C^{\Gamma}(u_{1},...,u_{d}):=\Gamma^{-1}(C(\Gamma(u_{1}),...,\Gamma(u_{d}))).
\end{equation*}

\noindent This transformation will affect the joint df $H$ and consequently
the df $G$ of the sum $Z.$ Their new forms will be denoted by $H^{\Gamma}$
and $G^{\Gamma}$ respectively. Morillas \cite{Mor05} describes some of the
existing families of distortion functions, among which the following are
frequently used:%
\begin{equation*}
\begin{array}{lcc}
\Gamma_{r} ( s ) =s^{r},\medskip & \text{for} & 0<r\leq1,\medskip \\ 
\Gamma_{\delta} ( s ) =\dfrac{\ln ( \delta s+1 ) }{\ln ( \delta+1 ) }%
,\medskip & \text{for} & \delta>0,\medskip \\ 
\Gamma_{\xi,\vartheta} ( s ) =\dfrac{ ( \xi+\vartheta ) s}{\xi s+\vartheta}%
,\medskip & \text{for} & \xi,\vartheta>0, \\ 
\Gamma_{\nu} ( s ) =\dfrac{s^{\nu}}{2-s^{\nu}}, & \text{for} & 0<\nu\leq1/3.%
\end{array}%
\end{equation*}
We call the corresponding distorted risk measures by \textit{copula
distorted risk measure} (CDRM) defined as%
\begin{equation*}
\pi_{\psi}^{\Gamma} [ Z ] =\int_{0}^{\infty}\psi ( 1-G^{\Gamma} ( z ) ) dz.
\end{equation*}
It is worth mentioning that if $X^{ ( 1 ) },...,X^{ ( d ) }$ are
independent, the corresponding copula function $C ( u_{1},...,u_{d} )
=\prod\limits_{i=1}^{d}u_{i}$ is called the product copula and denoted by $%
C^{\bot}.$ In this case, we have $C^{\Gamma}=C$ and therefore $%
\pi_{\psi}^{\Gamma} [ Z ] =\pi_{\psi} [ Z ] .$\medskip

\noindent The remainder of this paper is organized as follows. In Section 2,
we give a copula representation of the DRM's. In Section 3, we present a
more flexible class of copula given by the notion of distorted Archimedean
copulas. By the nice properties of this class and the copula representation
of the DRM, we introduce, in Section 4, the CDRM's. Finally, an illustrative
example, explaining the CDRM computation, is given in Section 5.

\section{\textbf{Copula representation of the DRM}}

\noindent Given a vector of risk losses $\mathbf{X}=(X^{(1)},...,X^{(d)}),$ $%
d\geq 2,$ with joint df $H$ and continuous margins $F_{i},$ $i=1,...,d.$ The
df of the rv $Z=\sum_{i=1}^{d}X_{i},$ is%
\begin{equation*}
G(t)=\int_{A(t)}dH(x_{1},..,x_{d}),\text{ for any }t\geq 0,
\end{equation*}%
where $A(t):=\{(x_{1},..,x_{d}):0\leq \sum_{i=1}^{d}x_{i}\leq t\}.$ Using
the representation $(\ref{eq5.3}),$ we get 
\begin{equation*}
G(t)=\int_{A(t)}dC(F_{1}(x_{2}),...,F_{d}(x_{d})).
\end{equation*}%
If we suppose that the copula $C$ and margins $F_{i}$ are differentiable
with densities $c$ and $f_{i},$ respectively, then%
\begin{equation*}
G(t)=\int_{A(t)}c(F_{1}(x_{1}),...,F_{d}(x_{d}))\prod%
\limits_{i=1}^{d}f_{i}(x_{i})dx_{1},...dx_{d}.
\end{equation*}%
The change of variables $F_{i}(x_{i})=u_{i},$ $i=1,...,d,$ yields%
\begin{equation}
G(t)=\int_{0}^{F_{d}(t)}\int_{0}^{F_{d-1}(t-F_{d}^{-1}(u_{d}))}...%
\int_{0}^{F_{1}(t-%
\sum_{i=0}^{d-2}F_{d-i}^{-1}(u_{d-i}))}c(u_{1},...,u_{d})du_{1}...du_{d}.
\label{eq5.6}
\end{equation}%
According to $(\ref{eq5.6}),$ the computation of the DRM corresponding to $%
Z, $ given in $(\ref{eq5.2}),$ requires the knowledge of the copula density
and the margins of vector $\mathbf{X}.$ In particular, for the bivariate
case $(d=2),$ we have%
\begin{equation*}
G(t)=\int_{0}^{F_{2}(t)}%
\int_{0}^{F_{1}(t-F_{2}^{-1}(u_{2}))}c(u_{1},u_{2})du_{1}du_{2}.
\end{equation*}%
Whenever $X_{1}$ and $X_{2}$ are independent, we have $c(u_{1},u_{2})=1,$
and therefore 
\begin{equation*}
G(t)=\int_{0}^{F_{2}(t)}F_{1}(t-F_{2}^{-1}(u_{2}))du_{2}=%
\int_{0}^{t}F_{1}(t-x)dF_{2}(x),
\end{equation*}%
which is the usual convolution of the $F_{i}$'s$.$

\section{\textbf{Distorted Archimedean copulas}}

\noindent In this paper, we focus on one important class of copulas called:
Archimedian copulas.\ This class contains several copula families useful in
dependence modelling. Their nice properties are captured by an additive
generator function $\varphi: [ 0,1 ] \rightarrow\lbrack0,\infty],$ which is
continuous, strictly decreasing and convex with $\varphi(1)=0.$ The main
advantage of the Archimedean copulas is the achievement of the reduction in
dimensionality of a $d$-variate distribution in a single argument. In
econometrics, this property has the potential to be of use in models of
limited dependent variables, especially those requiring some probabilistic
enumeration on high-dimensional subspaces. In the bivariate case, an
Archimedean copula is defined by%
\begin{equation*}
C(u,v)=\varphi^{ [ -1 ] } ( \varphi(u)+\varphi(v) ) ,
\end{equation*}
where%
\begin{equation*}
\varphi^{ [ -1 ] } ( t ) = \left\{ 
\begin{array}{ll}
\varphi^{-1} ( t ) , & 0\leq t\leq\varphi ( 0 ) ,\medskip \\ 
0, & \varphi ( 0 ) \leq t\leq\infty.%
\end{array}
\right.
\end{equation*}

\noindent Note that $\varphi^{ [ -1 ] }$ is continuous and non-increasing on 
$[0,\infty]$ and $\varphi$ is the unique generator up to a scaling constant.
If the terminal $\varphi(0)=\infty,$ the generator is called strict and $%
\varphi^{ [ -1 ] }=\varphi^{-1}.$ Numerous single-parameter families of
Archimedean copulas are listed in Table 4.1 in Nelsen \cite{Nel06}.
Particular examples are $\varphi_{\theta}(t)= ( t^{-\theta}-1 ) /\theta,$ $%
\varphi_{\alpha}(t)= ( -\ln t ) ^{\alpha}$ and $\varphi_{\beta} ( t ) =-\ln
( ( e^{-\beta t}-1 ) / ( e^{-\beta}-1 ) ) $ which are, respectively, the
generators of the Clayton family%
\begin{equation*}
C_{\theta}(u,v)= ( u^{-\theta}+v^{-\theta}-1 ) ^{-1/\theta},\text{ }%
\theta\geq0,
\end{equation*}
the Gumbel family%
\begin{equation*}
C_{\alpha}(u,v)=\exp \{ -[(-\ln u)^{\alpha}+(-\ln v)^{\alpha}]^{1/\alpha }
\} ,\text{ }\alpha\geq1,
\end{equation*}
and the Frank family%
\begin{equation*}
C_{\beta}(u,v)=-\frac{1}{\beta}\ln [ 1+\frac{ ( e^{\beta u}-1 ) ( e^{\beta
v}-1 ) }{e^{\beta}-1} ] ,\text{ }\beta\in\mathbb{R}\backslash \{ 0 \} .
\end{equation*}
The generators $\varphi_{\theta},$ $\varphi_{\alpha}$ and $\varphi_{\beta}$
are strict and therefore their corresponding copulas $C_{\theta},$ $%
C_{\alpha }$ and $C_{\beta}$ verify%
\begin{equation*}
C(u,v)=\varphi^{-1}(\varphi(u)+\varphi(v)).
\end{equation*}

\noindent Next, we discuss some properties of distortion functions acting on
bivariate Archimedean copulas. Given an Archimedean copula $C$ and a
strictly increasing bijection $\Gamma: [ 0,1 ] \mathbb{\ \rightarrow} [ 0,1
] ,$ we consider the function $C^{\Gamma}: [ 0,1 ] ^{2} \rightarrow [ 0,1 ] $
defined by%
\begin{equation*}
C^{\Gamma}(u,v)=\Gamma^{-1}(C(\Gamma(u),\Gamma(v))).
\end{equation*}
Under what conditions on $\Gamma,$ the function $C^{\Gamma}$ is an
Archimedean copula?\medskip

\noindent First, from Theorem 3.3.3. in Nelsen \cite{Nel06}, $C^{\Gamma}$ is
a copula if $\Gamma$ is concave and continuous on $[ 0,1 ] $ with $\Gamma (
0 ) =0$ and $\Gamma ( 1 ) =1.\ $The following Theorem gives an additional
condition so that the copula $C^{\Gamma}$ remains Archimedean. For
convenience, let $\mathbb{K}$ represents the set of the functions $\Gamma$
verifying the assumptions above.

\begin{theorem}
\label{Th5.1}Let $C$ be an Archimedean copula with generator $\varphi$ and
suppose that $\Gamma\in\mathbb{K},$ then the copula $C^{\Gamma}$ is
Archimedean if and only if $\varphi\circ\Gamma$ is convex.
\end{theorem}

\begin{proof}
Indeed, let $\varphi$ be the generator of the copula $C$ and let $\Gamma \in%
\mathbb{K},$ then%
\begin{equation*}
C^{\Gamma}(u_{1},...,u_{d})=\Gamma^{-1}(C(\Gamma(u_{1}),...,\Gamma(u_{d}))).
\end{equation*}
We have $\Gamma^{ [ -1 ] }=\Gamma^{-1},$ then%
\begin{equation*}
C^{\Gamma}(u_{1},...,u_{d})=\Gamma^{ [ -1 ] }\varphi^{ [ -1 ] } ( \varphi (
\Gamma ( u_{1} ) ) +...+\varphi ( \Gamma ( u_{d} ) ) ) .
\end{equation*}
It is easy to show that $\Gamma^{ [ -1 ] }\varphi^{ [ -1 ] }= (
\varphi\circ\Gamma ) ^{ [ -1 ] },$ it follows that%
\begin{equation}
C^{\Gamma}(u_{1},...,u_{d})=\mathcal{T}^{ [ -1 ] } ( \mathcal{T} ( u_{1} )
+...+\mathcal{T} ( u_{d} ) ) ,  \label{eq5.16}
\end{equation}
with $\mathcal{T}:=\varphi\circ\Gamma.$ From Theorem 4.1.4. Nelsen \cite%
{Nel06}, $C^{\Gamma}$ is Archimedean if and only if $\mathcal{T}$ is convex.
Notice that $\varphi\circ\Gamma$ is the generator of $C^{\Gamma}.$
\end{proof}

\begin{corollary}
The distortion function $t \rightarrow\Gamma^{\bot} ( t ) :=\exp ( -\varphi
( t ) ) $\ transforms any Archimedean copula $C$\ in the product copula $%
C^{\bot}.$
\end{corollary}

\begin{proof}
Straightforward.\medskip
\end{proof}

\noindent Next, we see the influence of the distortion of copulas on the
association measures. Kendall's tau and Spearman's rho are the most popular
measures of association, their representations in terms of the copula $C$
are given by%
\begin{equation*}
\tau=4\int_{0}^{1}\int_{0}^{1}C(u,v)dC(u,v)-1\text{ and }\rho=12\int_{0}
^{1}\int_{0}^{1} ( C(u,v)-uv ) dudv,
\end{equation*}
respectively. Let $\tau^{_{\Gamma}}$ and $\rho^{_{\Gamma}},$ respectively,
denote Kendall's tau and Spearman's rho of copula $C^{\Gamma}.$\ According
to Theorem 10\textbf{\ }in Durrleman \textit{et al.} \cite{Dur00}, we have
under suitable assumptions%
\begin{equation*}
1+\frac{\tau-1}{a^{2}}\leq\tau^{_{\Gamma}}\leq1+\frac{\tau-1}{b^{2}},
\end{equation*}
and%
\begin{equation*}
\frac{\rho+3}{a^{3}}-3\leq\rho^{_{\Gamma}}\leq\frac{\rho+3}{b^{3}}-3,
\end{equation*}
where $0<a\leq b<\infty$ are bounds for the derivative of $\Gamma.$

\section{\textbf{Risk measures for sums of losses}}

\noindent It may happen that the model (represented by the copula $C)$
chosen, to fit the data, does not provide enough information on the
dependence structure. This leads us to transform $C$ to a more flexible
copula $C^{\Gamma }$ of the same class. Consequently, the joint df of $%
\mathbf{X}$ may be represented, via Sklar's Theorem, as%
\begin{equation*}
H(x_{1},...,x_{d})=C^{\Gamma }(F_{1}(x_{1}),...,F_{d}(x_{d})).
\end{equation*}%
Suppose that $C$ is Archimedean with generator $\varphi ,$ then from Theorem %
\ref{Th5.1}, $C^{\Gamma }$ defined in (\ref{eq5.16}) is also Archimedean.\
Assume that $C^{\Gamma }$ has a density function $c^{\Gamma },$ then in view
of the representation (\ref{eq5.6}) the df $G^{\Gamma }$ of the sum $Z$ may
be written as%
\begin{equation*}
G^{\Gamma }(t)%
\begin{array}{c}
:=%
\end{array}%
\int_{0}^{F_{d}(t)}\int_{0}^{F_{d-1}(t-F_{d}^{-1}(u_{d}))}...%
\int_{0}^{F_{1}(t-\sum_{i=0}^{d-2}F_{d-i}^{-1}(u_{d-i}))}c_{\Gamma
}(u_{1},...,u_{d})du_{1}...du_{d}.
\end{equation*}%
Applying Wang's principle $(\ref{eq5.1})$ to the loss distribution $%
G^{\Gamma },$ we have%
\begin{equation*}
\pi _{\psi }^{\Gamma }[Z]:=\int_{0}^{+\infty }\psi (1-G^{\Gamma }(t))dt,
\end{equation*}%
which we call the CDRM. This may be considered as a manner of measuring the
risk $Z$ by distorting both the dependence structure and the distribution
tail, without losing the coherence feature. The CDRM adjusts the true
probability measure to give more weight to higher risk events and less
weight to the dependence structure. In other words, the simultaneous
transformations yield a new risk measure bounded by the expectation and
Wang's measure, that is 
\begin{equation}
\mathbb{E}[Z]\leq \pi _{\psi }^{\Gamma }[Z]\leq \pi _{\psi }[Z].
\label{eq5.21}
\end{equation}%
In the following example, we verify the previous inequalities on a selected
model.

\section{\textbf{Illustrative example}}

\noindent Let $X_{1}$ and $X_{2}$ be two risks with joint df represented by
the Clayton copula $C_{\theta },$ $\theta >0$ and Pareto-distributed margins 
$F_{1}$ and $F_{2}$ with respective parameters $0<\alpha _{1},\alpha _{2}<1,$
that is $F_{i}(x_{i})=1-x_{i}^{-1/\alpha _{i}},$ $x_{i}>1,$ $i=1,2.$
Kendall's tau of $C_{\theta }$ is $\tau =\theta /(\theta +2).$ Let $\psi
(x)=x^{1/\rho },$ $\rho \geq 1,$ and $\Gamma (t)=t^{1/\delta },$ $\delta
\geq 1.$ The distorted copula $C_{\theta }^{\Gamma },$ denoted by $C_{\theta
}^{\delta },$ is of Clayton type with generator $(\varphi \circ \Gamma
)(t)=(t^{-\theta /\delta }-1)/\theta $ and Kendall's tau is $\tau ^{\Gamma
}=(\theta /\delta )/(\theta /\delta +2).$ The df of the sum $Z=X_{1}+X_{2}$
is%
\begin{equation*}
G^{\delta }(t;\theta ,\alpha _{1},\alpha _{2})=\int_{1}^{1-t^{-1/\alpha
_{2}}}(\int_{1}^{1-(t-(1-v)^{-\alpha _{2}})^{-1/\alpha _{1}}}c_{\theta
}^{\delta }(u,v)du)dv,
\end{equation*}%
where 
\begin{equation*}
c_{\theta }^{\delta }(u,v)=(\theta /\delta +1)u^{-\theta /\delta
-1}v^{-\theta /\delta -1}(u^{-\theta /\delta }+v^{-\theta /\delta
}-1)^{-\delta /\theta -2},
\end{equation*}%
is the density of $C_{\theta }^{\delta }.\ $Figures 1 gives a preview of the
effect of the copula distortion.

\begin{center}
\begin{equation*}
\begin{tabular}{l}
\FRAME{itbpF}{2.6891in}{2.8588in}{0in}{}{}{claytond.eps}{\special{language
"Scientific Word";type "GRAPHIC";maintain-aspect-ratio TRUE;display
"USEDEF";valid_file "F";width 2.6891in;height 2.8588in;depth
0in;original-width 7.3929in;original-height 7.8607in;cropleft "0";croptop
"1";cropright "1";cropbottom "0";filename 'Articles tex/Article Brahimi et
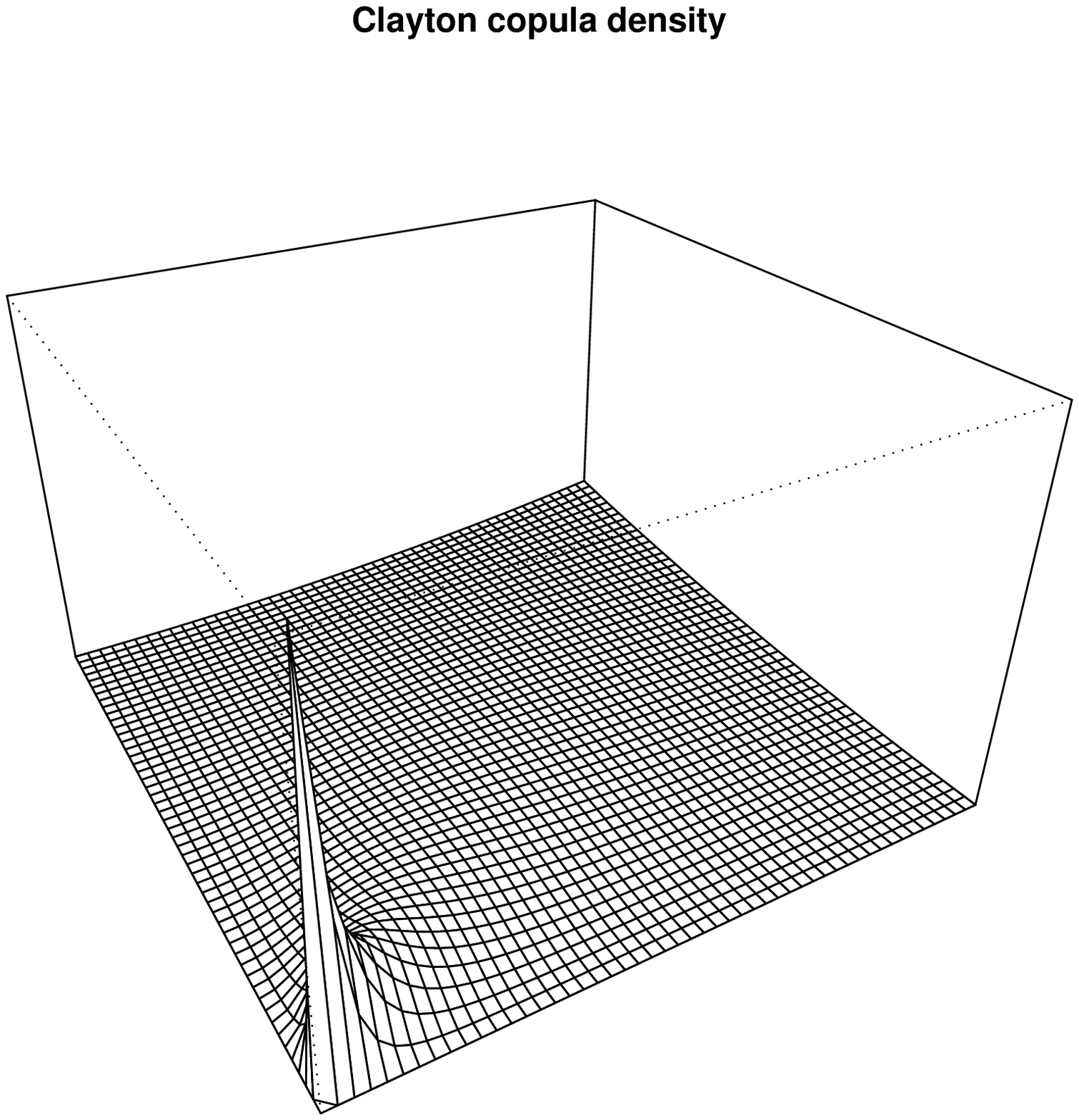';file-properties "XNPEU";}}\FRAME{itbpF}{2.6896in}{2.8573in}{%
0in}{}{}{distortedcop.eps}{\special{language "Scientific Word";type
"GRAPHIC";maintain-aspect-ratio TRUE;display "USEDEF";valid_file "F";width
2.6896in;height 2.8573in;depth 0in;original-width 7.3929in;original-height
7.8607in;cropleft "0";croptop "1";cropright "1";cropbottom "0";filename
'Articles tex/Article Brahimi et 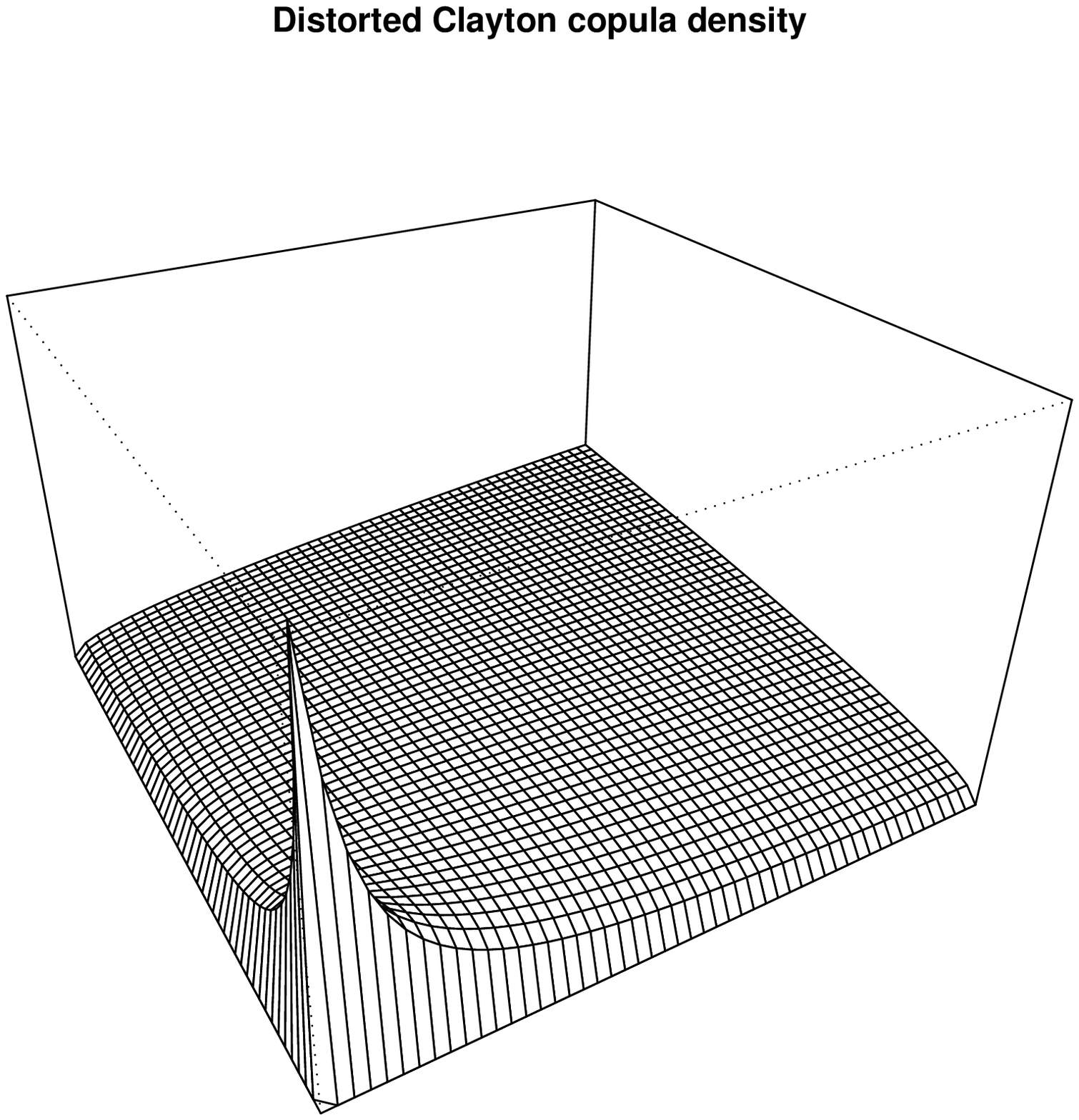';file-properties
"XNPEU";}} \\ 
\multicolumn{1}{c}{Figure 1. Clayton copula density with $\theta =2$ (left
panel)} \\ 
\multicolumn{1}{c}{and its distorted copula density with $\delta =4$ (right
panel)}%
\end{tabular}%
\ \ \ 
\end{equation*}%
\bigskip
\end{center}

\noindent The DRM and the CDRM of $Z$ are respectively denoted by 
\begin{equation*}
\pi _{\rho }[Z]=\int_{2}^{\infty }(1-G(t))^{1/\rho }dt,\quad \text{and}\quad
\pi _{\rho }^{\delta }[Z]=\int_{2}^{+\infty }(1-G^{\delta }(t))^{1/\rho }dt.
\end{equation*}%
We select a Pareto model with $\theta =3/2,$ $\alpha _{1}=1/3$ and $\alpha
_{2}=1/5.$ We obtain $\mathbb{E}(Z)=0.750$ and $\tau =0.428.$ For two
different tail distortion parameters $\rho =1.2$ and $\rho =1.4$ the
respective DRM's are $1.225$ and $2.091.$\ The CDRM's for distinct values of
the copula distortion parameter $\delta $ are summarized in Tables 1 and 2,
where we see that the inequalities $(\ref{eq5.21})$ are satisfied for any
value of the copula distortion parameter. This is well shown graphically in
Figure 2 in which the three risk measures of $(\ref{eq5.21})$ are plotted as
functions of $\delta .$

\begin{center}
$%
\begin{tabular}{l|ccccccccc}
\hline
$\delta$ & ${\small 1}$ & ${\small 1.5}$ & ${\small 2}$ & ${\small 2.5}$ & $%
{\small 3}$ & ${\small 3.5}$ & ${\small 4}$ & ${\small 5}$ & ${\small 6}$ \\ 
\hline\hline
$\tau^{\delta}$ & ${\small 0.428}$ & ${\small 0.333}$ & ${\small 0.272}$ & $%
{\small 0.230}$ & ${\small 0.200}$ & ${\small 0.176}$ & ${\small 0.157}$ & $%
{\small 0.130}$ & ${\small 0.111}$ \\ 
$\pi_{\rho}^{\delta} [ Z ] $ & ${\small 1.225}$ & ${\small 1.030}$ & $%
{\small 0.988}$ & ${\small 0.969}$ & ${\small 0.964}$ & ${\small 0.961}$ & $%
{\small 0.958}$ & ${\small 0.953}$ & ${\small 0.950}$ \\ \hline\hline
\multicolumn{10}{c}{} \\ 
\multicolumn{10}{l}{Table 1.\textbf{\ }CDRM's and transformed Kendall tau of
the sum of two Pareto-distributed} \\ 
\multicolumn{10}{l}{risks with tail distortion parameter $\rho=1.2.$}%
\end{tabular}
$\medskip

$%
\begin{tabular}{l|ccccccccc}
\hline
$\delta$ & ${\small 1}$ & ${\small 1.5}$ & ${\small 2}$ & ${\small 2.5}$ & $%
{\small 3}$ & ${\small 3.5}$ & ${\small 4}$ & ${\small 5}$ & ${\small 6}$ \\ 
\hline\hline
$\tau^{\delta}$ & ${\small 0.428}$ & ${\small 0.333}$ & ${\small 0.272}$ & $%
{\small 0.230}$ & ${\small 0.200}$ & ${\small 0.176}$ & ${\small 0.157}$ & $%
{\small 0.130}$ & ${\small 0.111}$ \\ 
$\pi_{\rho}^{\delta} [ Z ] $ & ${\small 2.091}$ & ${\small 1.801}$ & $%
{\small 1.736}$ & ${\small 1.712}$ & ${\small 1.703}$ & ${\small 1.699}$ & $%
{\small 1.694}$ & ${\small 1.685}$ & ${\small 1.680}$ \\ \hline\hline
\multicolumn{10}{c}{} \\ 
\multicolumn{10}{l}{Table 2\textbf{. }CDRM's and transformed Kendall tau of
the sum of two Pareto-distributed} \\ 
\multicolumn{10}{l}{risks with tail distortion parameter $\rho=1.4.$}%
\end{tabular}
$

\begin{equation*}
\begin{array}{c}
\FRAME{itbpF}{3.0493in}{3.0493in}{0in}{}{}{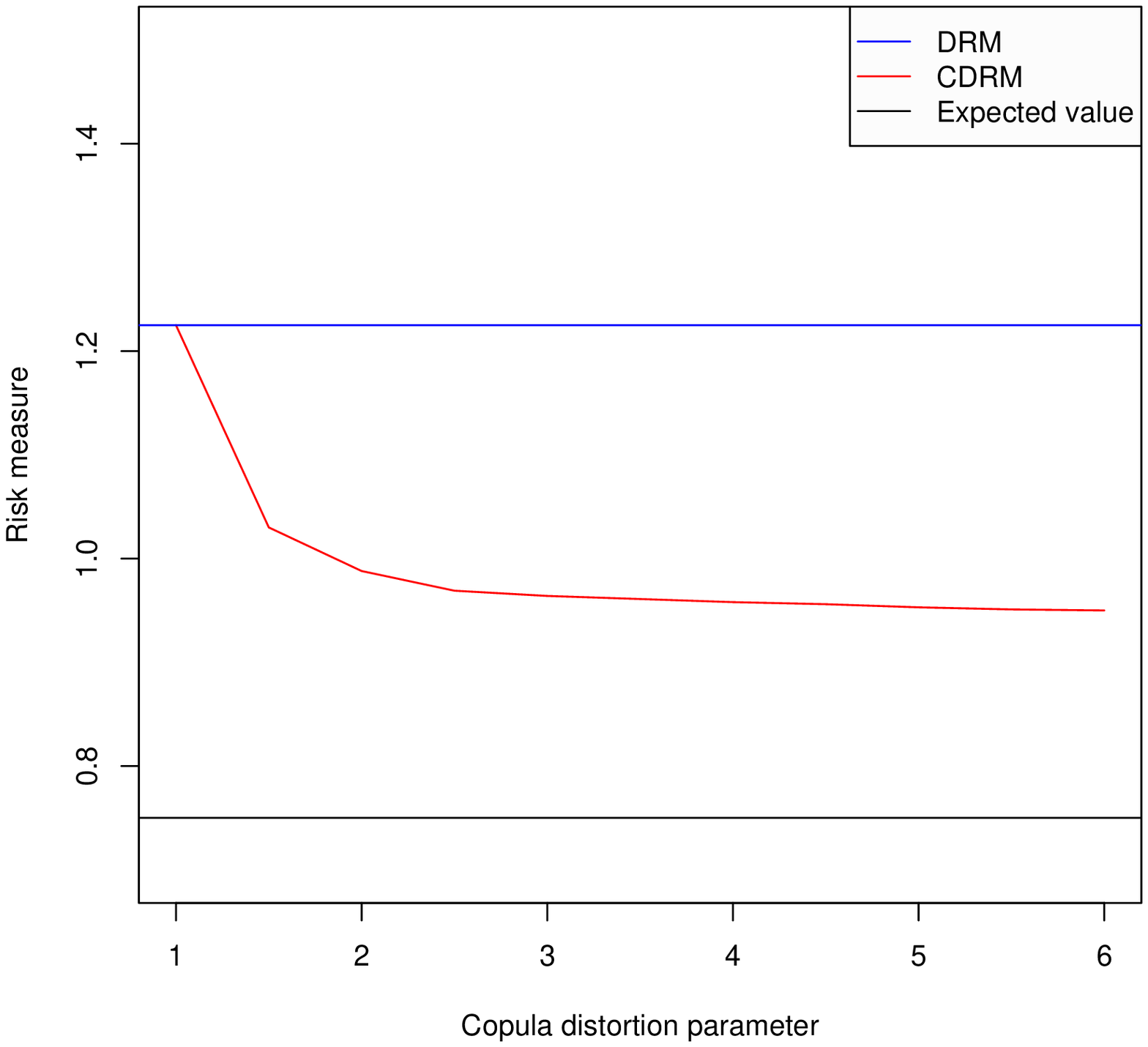}{\special{language
"Scientific Word";type "GRAPHIC";maintain-aspect-ratio TRUE;display
"USEDEF";valid_file "F";width 3.0493in;height 3.0493in;depth
0in;original-width 6.9998in;original-height 6.9998in;cropleft "0";croptop
"1";cropright "1";cropbottom "0";filename '1.eps';file-properties "XNPEU";}}
\\ 
\multicolumn{1}{l}{\text{Figure 2. Risk measures of the sum of two
Pareto-distributed risks}} \\ 
\multicolumn{1}{l}{\text{with tail distortion parameter }\rho =1.2.}%
\end{array}%
\end{equation*}
\end{center}

\noindent Taking $\delta =1$\ means that\ no distortion on the dependence
structure is made, that is $C^{1}=C,$\ and $\pi _{\rho }^{1}[Z]=\pi _{\rho
}[Z].$ In other words, the CDRM with $\delta =1$ reduces to Wang's DRM,
which can be seen in the second columns of Tables 1 and 2. This fact is also
clear in Figure 2. On the other hand, as $\delta $ increases, the
transformed Kendall's tau decreases meaning that the dependence gets weaker
(see the second lines of Tables 1 and 2). Moreover, starting from some $%
\delta $ the CDRM values become roughly constant while being always greater
than the expectation (see the third lines of Tables 1 and 2).

\section{\textbf{Concluding remarks}}

\noindent In portfolio analysis, the dependence structure has a major role
to play when quantifying risks. This led us to think of risk measure taking
into account this fact in addition to the tail behavior. In this paper, we
proposed a risk measure for the sum of two dependent losses by
simultaneously transforming the distribution tail and the copula, which
represents the dependence between the margins, by means of two distortion
functions.\ We obtained a coherent measure that we called the \textit{Copula
Distorted Risk Measure.\ }This new measure has the characteristic to be
greater than the expectation and less than the popular Wang's DRM. In the
insurance business, the main advantage of this property is to reduce Wang's
premium while respecting the standard axioms of the premium
principle.\bigskip

\noindent \textbf{Acknowledgment}\newline
\noindent The authors would like to thank the Editors and the anonymous
reviewers for their comments and suggestions that improved the quality of
the paper.


\begin{thebibliography}{99}
\bibitem{Art99} Artzner, P., Delbaen, F., Eber, J.M., Heath, D., 1999.
Coherent measures of risk. \textit{Mathematical Finance} \textbf{9}, 203-228.

\bibitem{Bur06} Burgert, C., R\"{u}schendorf, L., 2006. Consistent risk
measures for portfolio vectors. \textit{Insurance Math. Econom.} \textbf{38}%
, 289-297.

\bibitem{Cra08} Crane, G., van der Hoek, J., 2008. Using distortions of
copulas to price synthetic CDOs. \textit{Insurance Math. Econom.} \textbf{42}%
, 903-908.

\bibitem{Dar04} Darkiewicz, G., Dhaene, J., Goovaerts, M.J., 2004.
Distortion risk measures for sums of random variables. Bl\"{a}tter der
DGVFM, \textit{Springer Berlin/Heidelberg} \textbf{26}, 631-641.

\bibitem{Den94} Denneberg, D., 1994. Non-additive measure and integral.
Theory and Decision Library 27, Kluwer Academic Publilshers.

\bibitem{Den05} Denuit, M., Dhaene, J., Goovaerts, M. and Kaas, R., 2005. 
\textbf{Actuarial Theory for Dependent Risks : Measures, Orders and Models}.
Jhon Wiley \& Sons, Ltd.

\bibitem{Dur00} Durrleman, V., Nikeghbali, A., Roncalli, T., 2000. A simple
transformation of copulas. Groupe de Recherche Op\'{e}rationnelle, Cr\'{e}%
dit Lyonnais, France.

\bibitem{Fre98} Frees, W.E., Valdez, E.A., 1998. Understanding relationships
using copulas. \textit{N. Am. Actuar. J.} \textbf{2}, 1-25.

\bibitem{Fur08a} Furman, E., Zitikis, R., 2008a. Weighted premium
calculation principles. \textit{Insurance Math. Econom.} \textbf{42},
459-465.

\bibitem{Fur08b} Furman, E., Zitikis, R., 2008b. Weighted risk capital
allocations. \textit{Insurance Math. Econom.} \textbf{43}, 263--269.

\bibitem{Gen01} Genest, C., Rivest, L.P., 2001. On the multivariate
probability integral transformation. \textit{Statist. Probab. Lett.} \textbf{%
53}, 391-399.

\bibitem{Hur98} H\"{u}rlimann, W., 1998. On stop-loss order and the
distortion pricing principle. \textit{ASTIN Bulletin} \textbf{28}, 119-134.

\bibitem{Mor05} Morillas, P.M., 2005. A method to obtain new copulas from a
given one. \textit{Metrika} \textbf{61}, 169-184.

\bibitem{Nel06} Nelsen, R., 2006. \textbf{An introduction to copulas}.
Springer Verlag, New York.

\bibitem{Skl59} Sklar, A., 1959. Fonctions de r\'{e}partition \`{a} n
dimensions et leurs marges. \textit{Inst. Statist. Univ. Paris } \textbf{8},
229-231.

\bibitem{Val10} Valdez, E.A., Xiao, Y., 2010. On the distortion of a copula
and its margins. \textit{Scand. Actuar. J.} (in press).

\bibitem{Wan95} Wang, S.S., 1995. Insurance pricing and increased limits
ratemaking by proportional hazards transforms. \textit{Insurance Math.
Econom.}, \textbf{17}, 43-54.

\bibitem{Wan00} Wang, S.S., 2000.\ A class of distortion operators for
pricing financial and insurance risks. \textit{Journal of Risk and Insurance}
\textbf{67}, 15-36.
\end{thebibliography}
\end{document}